# Magnetoelectric Wireless Power Transfer Receivers for Sub-Millimeter Scale Implantable Bioelectronics

Authors: Jonathan Tan[1], Serene Feng[1], Thomas Mion[2], Sydney Acosta[1], Konrad Bussmann[2], Peter Finkel[2], Roy Olsson[1*]

[1]Department of Electrical and Systems Engineering, University of Pennsylvania, Philadelphia, PA, USA

[2]Materials Science and Technology Division, US Naval Research Laboratory, Washington, DC 02375, USA

*Email: rolsson@seas.upenn.edu

Abstract

Implanted biomedical devices for continuous physiological monitoring require reliable wireless power delivery to anatomically constrained and sensitive sites where batteries are impractical or hazardous. Existing wireless power transfer (WPT) modalities, including inductive coils and ultrasonic transducers, suffer from performance degradation as receiver dimensions are miniaturized, limiting their application for space-limited locations such as fingertips, ocular cavity, and intracranial sites. We present a microelectromechanical systems (MEMS) magnetoelectric WPT receiver integrating piezoelectric aluminum scandium nitride and magnetostrictive iron-cobalt/silver multilayer thin-films, microfabricated within a footprint of 770 × 360 × 2.25 μm. The receiver operates at its 100 kHz flexural-mode resonance with a quality factor of 1180 and delivers 1.0 μW to a matched 10 kΩ load under a standard-compliant 205 $\mu T_{RMS}$ excitation field. This device exhibits a magnetoelectric coupling coefficient of 1274 V $cm^{-1}$ $Oe^{-1}$ and a power figure of merit ($FOM_P$) of 4.05 × $10^{-3}$ mW $cm^{-3}$ $Oe^{-2}$ $Hz^{-1}$, both the highest reported values of comparable magnetoelectric WPT receivers. The FOM is 20-fold higher than the next best comparable-frequency device. These miniaturized devices establish MEMS magnetoelectric transduction as a scalable platform for battery-free wireless powering of microwatt-class implants in the most anatomically demanding regions of the body.

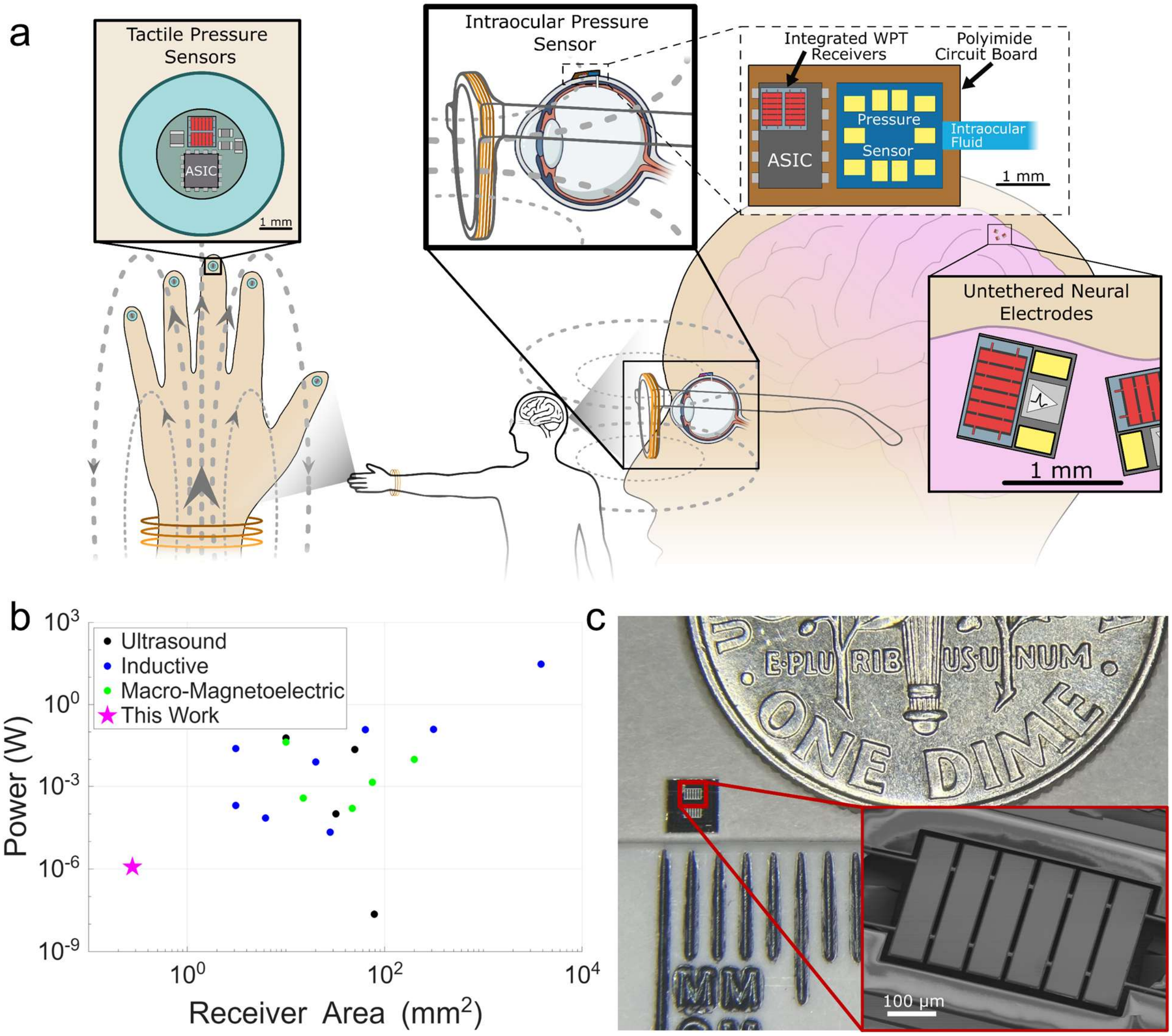

Figure 1. MEMS magnetoelectric WPT for microwatt implants. a) A system concept of the WPT receivers powering force sensors in the fingers[1], an intraocular pressure sensor[2], and tetherless neural electrodes[3]. b) Output power and size comparison of the MEMS WPT receiver with inductive, ultrasonic, and magnetoelectric WPT receivers[4-19]. c) Size of two WPT receivers on a silicon die compared to a US dime, and a scanning electron micrograph of the WPT receiver, showing the released resonant structure.

## Introduction

The biomedical internet-of-things envisions a future of wide-ranging implantable and wearable medical devices for continuous and remote measurement of biomarkers, providing valuable data for healthcare professionals to better diagnose and treat conditions[20, 21]. Commercially available medical devices such as pacemakers and continuous glucose monitors rely on an internal battery for power. These batteries can fail or require replacement over time, necessitating invasive surgeries[22, 23]. Biomedical devices implanted into anatomically constrained and sensitive sites, including those for peripheral nerve recording[24-26], brain neurotransmitter sensing[27], electrocardiography[28], blood flow monitoring[29], pressure sensing in the fingers[1], and intraocular and intracranial pressure[2, 30], all require power for operation while being severely restricted in overall system volume. For these devices, implanting them with a bulky battery is infeasible, as the proximity to sensitive organs limits the space available for implants.

Wireless power transfer (WPT) offers the ability to continuously power implanted, battery-free medical devices through an external stimulus[31, 32]. In anatomically sensitive locations such as the head and torso, the permitted incident power transmitted to the body is frequency-dependent and highly limited, necessitating thoughtful receiver design[33]. WPT receivers for implantation into space-constrained and sensitive sites should therefore exhibit high power density, through both minimal sizing and high coupling between the external stimulus power and the delivered electrical power. WPT receivers used in research and clinical bioelectronics operate based on several modalities, including inductive coupling, ultrasonic acoustics, and the magnetoelectric (ME) effect[31, 32].

Inductively coupled coils operate on the principle of electromagnetic induction, generating electrical current in a receiving coil through a time-varying magnetic field produced by a transmit coil. As the receiver dimensions decrease, the magnetic flux through the coil decreases, directly impacting power delivered to the load[32]. Additionally, inductive coils are inherently sensitive to misalignment and typically operate at MHz frequencies, at which the permissible whole-body magnetic field strength is significantly lower than at frequencies below 100 kHz, described by IEEE Std C95.1-2019[33, 34]. Although operating at lower frequencies enables higher magnetic-field strengths, it incurs a separate penalty of decreased coil quality factor ($Q = \omega L/R$)[35]. Together, these constraints make inductive coils poorly suited for implanted devices in space-constrained locations where magnetic field strengths are highly limited.

Ultrasonic WPT receivers offer an alternative approach but introduce a distinct set of scaling limitations. Ultrasonic WPT receivers typically employ a resonant piezoelectric plate or diaphragm that generates a voltage upon displacement caused by acoustic waves generated outside the body[36]. As the dimensions decrease, the resonant frequency of the ultrasonic transducer increases. At these higher frequencies, biological tissues become increasingly absorptive to ultrasound energy[37]. To counteract the losses introduced by the absorptive tissue medium, higher incident power at the surface of the skin is required. The limits to diagnostic ultrasound power set by the U.S. Food and Drug Administration (FDA) scale as $f^{-1/2}$, with excessive acoustic beam pressure risking cavitation in tissue[36]. Miniaturization of ultrasonic WPT transducers therefore simultaneously increases the incident power required to overcome tissue losses and reduces the maximum power permitted by FDA safety limits.

The scaling limitations of WPT through inductive coupling and ultrasonic transduction motivate a magnetoelectric WPT approach, an alternative modality that operates with negligible losses to the body and without the performance penalties that accompany miniaturization of both inductive and ultrasonic WPT. ME WPT receivers operate through the strain-coupling of piezoelectric and magnetostrictive materials. In an external time-varying magnetic field, strain generated in the magnetostrictive layer is mechanically coupled to the piezoelectric layer, producing a time-varying voltage and delivering power to a load[38]. At low frequencies, magnetic field penetration through tissue is lossless and the permissible magnetic field strengths are high[33], allowing ME WPT receivers to avoid the fundamental scaling limitations that hamper inductive and ultrasonic WPT modalities at small dimensions.

Piezoelectric materials transduce energy between mechanical and electrical domains, generating an electric field in response to applied stress. This response is described by $D_i = d_{ij}T_j$, where $D_i$ is the electric displacement field in the i direction, $d_{ij}$ is the piezoelectric coefficient, and $T_j$ is the stress applied in the j direction; materials with larger piezoelectric coefficient produce a larger electric field for a given mechanical strain. Lead zirconate titanate (PZT) is a common piezoelectric material in ME WPT devices because of its high piezoelectric coefficient[39-41]. Aluminum nitride (AlN) is common in micrometer-scale sensors and devices because of its intrinsic piezoelectricity and CMOS back-end-of-line compatibility[42-44]. Scandium substitution into the AlN lattice (AlScN) significantly enhances the piezoelectric coefficient ($d_{31}$) relative to pure AlN[45, 46]. Unlike PZT, AlScN is lead-free, making it more suitable for implantable applications by eliminating biocompatibility risks and bulky enclosures required when implanting PZT devices.[38, 41, 47]. AlScN is therefore used as the piezoelectric layer in this work.

The performance of a ME WPT receiver depends equally on the properties of the magnetostrictive layer, which transduces energy between the applied AC magnetic field and mechanical strain. The piezomagnetic coefficient, $d_{33,m} = d\lambda/dH$, describes this response, the magnetostrictive strain ($\lambda$) generated per unit of applied magnetic field ($H$), for a magnetic field and strain in the same direction[38]. The $d_{33,m}$ of a magnetostrictive material is dependent on the DC magnetic field across the material; as such, a static magnetic field may be required to optimally bias the magnetostrictive layer to provide the largest $d_{33,m}$ response[38, 40]. In WPT applications, minimized optimal bias fields are desired, as the need for DC magnetic fields requires permanent magnets or electromagnets. Common magnetostrictive materials for WPT devices include iron-cobalt (FeCo) alloys[42], FeGaB[44], and Metglas[39-41]. Multilayers of $Fe_{50}Co_{50}$ and silver ($Fe_{50}Co_{50}$/Ag ML) have been shown to reduce the coercive field and increase the piezomagnetic coefficient of the magnetostrictive film, as compared to a single layer of $Fe_{50}Co_{50}$[48]. Thin-film $Fe_{50}Co_{50}$/Ag ML is therefore used as the magnetostrictive material in this work.

While centimeter and millimeter-scale ME WPT have demonstrated milliwatt-level power delivery[39, 49, 50], devices implanted in the most anatomically sensitive and constrained sites, such as neurotransmitter sensing in the brain[27] and pressure sensing in the fingertips[1] or eye[2], require at least an order of magnitude decrease in WPT receiver size. Microelectromechanical systems (MEMS), micro-scale devices that transform energy between electrical and mechanical domains, provide a path to this miniaturization, enabling ME WPT receivers with micrometer-scale dimensions through photolithography, deposition, and etching microfabrication techniques. Critically, high strain-coupling between magnetostrictive and piezoelectric materials can be achieved through the deposition of these materials directly onto one another without the use of mechanically lossy polymer adhesives used in larger ME WPT receivers[9, 50, 51]. Operation of a MEMS ME WPT receiver at its mechanical resonance amplifies the device velocity by its quality factor Q, describing the ratio of mechanical energy stored to the energy lost per cycle of its resonance. Output power delivered to a matched load increases linearly with the quality factor of the resonance, making Q a critical factor in the design of ME WPT receivers. The MEMS ME WPT receiver in this work demonstrates significant scaling benefits over inductive, ultrasonic, and macro-scale magnetoelectric WPT receivers in literature, as shown in Figure 1b.

In this work, we present a MEMS ME WPT receiver with sub-millimeter dimensions of 770 × 360 × 2.25 μm, designed to operate at its 100 kHz flexural-mode resonance within the IEEE C95.1-2019 whole-body magnetic field safety limit[33]. The WPT receiver is able to deliver 1 μW of power to a matched resistive load, achieving record magnetoelectric coupling coefficient and power figure of merit among comparable-frequency ME WPT receivers. With microwatt output power and sub-millimeter dimensions, this device is well suited to power sensors and bioelectronics in anatomically constrained sites including intraocular and fingertip pressure sensors[1, 2], and neural recorders and readout electronics[25, 52], as illustrated in Figure 1a. The results presented here establish MEMS ME transduction as a platform for battery-free wireless powering of implants in the most anatomically demanding and sensitive regions of the body.

Results

Flexural-Mode Resonator

The MEMS ME WPT device was designed around the resonant frequency and allowable incident magnetic field at the skin. ME WPT receiver output power increases linearly with frequency and with the square of the excitation AC magnetic field[53], making the choice of operating frequency critical toward maximizing the power delivery of the device at safe magnetic field strengths. Magnetically actuating ME WPT receivers at 100 kHz takes advantage of the flat maximum whole-body magnetic field limit below 100 kHz before a decrease with 1/f above 100 kHz[33]. Operation at 100 kHz maximizes the product of permissible squared magnetic field and frequency, as shown in Supplementary Note 1.

Achieving a mechanical resonance at 100 kHz in a sub-millimeter device requires careful resonant mode selection, as miniaturization generally increases resonant frequency. The flexural bending-mode resonance was chosen for its out-of-plane compliance, which yields a significantly lower resonant frequency than length- or width-extensional resonant modes of equivalent resonator dimensions[54]. A MEMS free-free beam geometry was designed to take advantage of this compliance, achieving approximately 100 kHz flexural-mode resonance while maintaining sub-millimeter dimensions.

The flexural-mode resonant WPT receiver consists of six free-free beams, each 360 × 90 μm, connected by 10 × 10 μm tethers, creating a total resonant structure dimension of 360 × 590 μm. Tethers of dimensions 10 × 90 μm anchor the free-free beams to the substrate at the quasi-nodal points of the flexural resonant mode, approximately halfway between the center and edge of the beam[55]. Anchoring at the quasi-nodal points minimizes anchor losses that dampen the motion of the resonating beams. A comparison between the size of the WPT device with a U.S. dime along with a scanning electron micrograph of the resonant structure is shown in Figure 1c. The material stack of the device from the intrinsic Si substrate includes 15 nm Ti / 150 nm Pt / 1 μm AlScN / 10 nm Ti / 100 nm Pt / 1 μm $Fe_{50}Co_{50}$/Ag ML, with the $Fe_{50}Co_{50}$/Ag multilayer only on the resonant beam area. The microfabrication flow is described in the methods section, with the cross-section fabrication flow depicted in Supplementary Note 5. The simulated flexural mode shape and surface stress distribution from a COMSOL eigenfrequency study are shown in Figure 2a, and an optical micrograph of the fabricated receiver showing the suspended resonant structure following isotropic $XeF_2$ silicon release is shown in Figure 2b.

The resonance of the WPT receiver was characterized by measuring the two-port $S_{21}$ transmission S-parameter under 4 mBar vacuum using a network analyzer (Agilent Technologies E5061B). A modified-Butterworth van Dyke (mBVD) model was fit to the $S_{21}$ measurement to extract the device parameters. The mBVD circuit model with fitted parameters is shown in Supplementary Note 2. The full S parameters were converted to admittance Y parameters, with the $Y_{21}$ admittance plotted in Figure 2c along with the admittance of the fit mBVD model. The WPT receiver exhibits a resonant frequency of 98.825 kHz, a quality factor of 1180, an electromechanical coupling coefficient, $k_{eff}^2$, of 0.55%, and an admittance of -79.89 dB, corresponding to a series resistance at resonance of 9.87 kΩ.

The ME output of the receiver is partially governed by the piezomagnetic coefficient, $d_{33,m}$, of the magnetostrictive $Fe_{50}Co_{50}$/Ag ML, which is maximized at an optimal DC bias magnetic field. To identify this optimal bias, a DC magnetic field was swept using wound electromagnets while applying a 3.96 $\mu T_{RMS}$ AC magnetic field at the 98.8 kHz device resonance and measuring the device output into a 50 Ω power spectrum analyzer (full testing setup described in Methods). The magnetoelectric coupling coefficient, $\alpha_{ME} = \partial E/\partial H = V_{OC} t_{PE}^{-1} H_{AC}^{-1}$, describing the relationship between induced electric field across the piezoelectric layer and the applied AC magnetic field, was calculated from the measured output power and is plotted as a function of DC bias field in Figure 2d. The response exhibits a local minimum at 2 mT due to intrinsic hysteretic effects in the magnetostriction response, combined with residual stress in the magnetic layer[56]. The maximum of the magnetostriction response to the AC H-field was observed when the device was biased around 3.2 mT. However, upon saturating the magnetostrictive film at 6 mT and returning the bias field to lower magnitudes, the magnetostriction response increased and the optimal bias field shifted to 4 mT, as shown in Figure 2d. Unless otherwise stated, all experiments were performed after saturating the magnetic field at 6 mT and subsequently maintaining the bias field at 4 mT to optimally bias the magnetostrictive film.

The ME coupling coefficient was additionally determined from the open-circuit voltage of the WPT receiver. One port of the WPT receiver was connected to a SMA short standard, with the other port connected to the 1 MΩ / 15 pF input of a voltage spectrum analyzer (Stanford Research Systems SR770). The vacuum chamber was pumped to 4 mBar vacuum and 4 mT bias magnetic field was applied to the device. An AC magnetic field with magnitude of 0.396 $\mu T_{rms}$ was applied, and 0.464 $mV_{RMS}$ was measured in the spectrum analyzer. Correcting for the loading effect of the 15 pF input capacitance of the spectrum analyzer, the open-circuit voltage is calculated to be 0.5050 $mV_{RMS}$, yielding a ME coupling coefficient, $\alpha_{ME}$, of 1274 V $cm^{-1}Oe^{-1}$ for the 1 µm thick AlScN piezoelectric layer. This represents the highest ME coupling coefficient reported among ME devices operating within an order of magnitude of this frequency[38, 42].

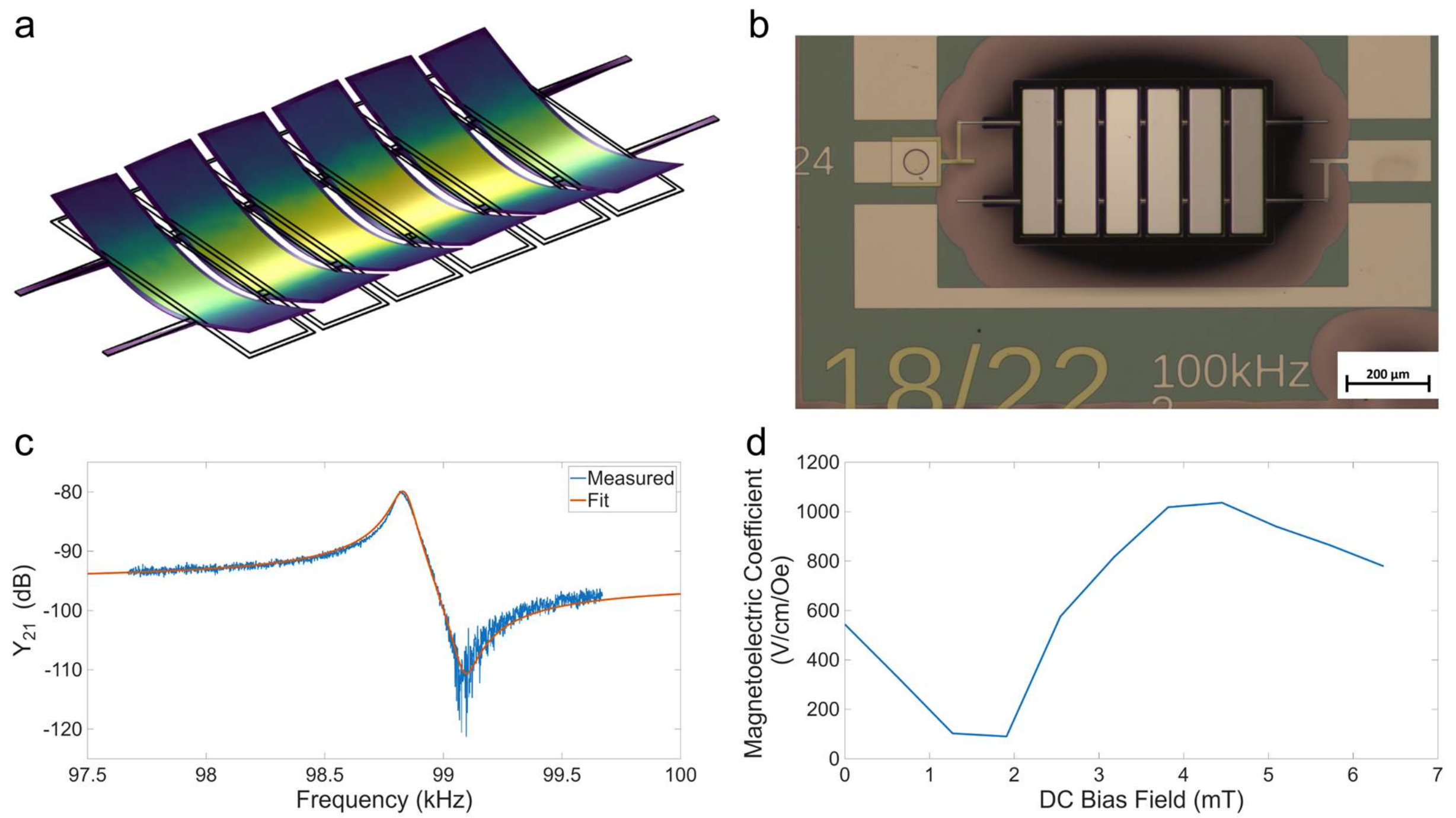

Figure 2. WPT receiver resonance and magnetic biasing. a) Modal FEM simulation of 100 kHz flexural mode resonance of the WPT receiver, with stress plotted on the device surface. b) Optical micrograph of WPT receiver, showing a stress-neutral released resonator. c) $Y_{21}$ admittance curve of measurement and mBVD model fit of the WPT receiver. d) ME coupling coefficient across various DC magnetic field bias strengths.

Wireless Power Transfer

While a MEMS ME device of comparable size has been reported in the literature, only ME coupling coefficients were reported[42]. However, power delivery to a matched load under realistic magnetic field conditions is the measure that directly determines the ability of a receiver to power implanted bioelectronics. To demonstrate this, the WPT receiver was driven at magnetic field strengths up to the 205 $\mu T_{RMS}$, the whole-body limit of IEEE C95.1-2019[33], with power delivered to a matched resistive load.

The optimal load impedance for maximum power transfer was determined from the mBVD circuit model fit to the $S_{21}$ measurement (Supplementary Note 2). At resonance, the impedances of the motional inductor and capacitor cancel, leaving the motional resistance as the sole impedance in the motional branch. Since the electrical parallel capacitance impedance is much greater than the motional resistance at the 100 kHz resonance, the total device impedance at resonance is approximately equal to the motional resistance alone. The motional resistance of approximately 10 kΩ, determined from the -39.9 dB insertion loss of the $S_{21}$ measurement (Supplementary Note 3), sets the optimal matched load impedance for maximum power delivery. A 10 kΩ resistor was connected across the output port of the WPT receiver to ground, with a high-impedance voltage spectrum analyzer connected in parallel to monitor the voltage drop across the load. A time-varying excitation magnetic field was applied at frequencies swept through the device resonance, with the current through the AC magnetic field coils monitored to determine the applied magnetic field strength. The power delivered by the WPT receiver was calculated from the RMS voltage across the load resistor as $P = V_{RMS}^2 / R_L$. The WPT receiver delivered power to the matched 10 kΩ load across a range of magnetic field strengths spanning below and slightly above the 205 $\mu T_{RMS}$ whole-body limit of IEEE C95.1[33], reaching an output of 1.03 μW at 207.95 $\mu T_{RMS}$ and 1.18 μW at the highest tested field strength of 215.6 $\mu T_{RMS}$. Scaling the 1.03 μW output power at 207.95 $\mu T_{RMS}$ to 205 $\mu T_{RMS}$ by its squared magnetic field relation yields an output power of 1 μW at the IEEE Std C95.1-2019 limit.

At high magnetic field strengths, the resonant response exhibits a third-order Duffing nonlinearity, seen as an asymmetric frequency response around the resonant peak, shown in Figure 3a. This Duffing nonlinearity is caused by mechanical spring softening of the device at large displacements. At greater magnetic field strengths, the nonlinearity causes a discontinuity in the frequency response (Supplementary Note 4). Despite this nonlinearity, the output power is consistent with the expected squared magnetic field dependence across the full range of magnetic field strengths, with no compression of the device response observed at magnetic fields up to 215.6 $\mu T_{RMS}$.

For meaningful comparison with existing ME WPT receivers, the power figure of merit ($FOM_p$) proposed by Saha et al.[53] is used:

$$FOM_P = \frac{P}{vol\ H_{ac}^2 f}$$

where P is measured output power, vol is the device volume, $H_{ac}$ is the applied magnetic field, and f is the operating frequency. This FOM normalizes output power by device volume, magnetic field strength, and frequency, providing a more rigorous basis for comparison of ME WPT receivers than the ME coupling coefficient, since the FOM directly reflects power delivery performance.

The MEMS ME WPT receiver presented achieves a $FOM_P$ of $4.05 \times 10^{-3}$ mW cm$^{-3}$ Oe$^{-2}$ Hz$^{-1}$, taking the device volume as the smallest rectangular prism encompassing the resonator and tethers, or $5.05 \times 10^{-3}$ mW cm$^{-3}$ Oe$^{-2}$ Hz$^{-1}$ considering the resonator volume alone. The corresponding power density is 400 mW cm$^{-3}$ Oe$^{-2}$ for the larger volume. A comparison of the $FOM_P$ and power density with other ME WPT receivers reported in literature are shown in Figures 3c and 3d respectively. The comparison plot of $FOM_P$ additionally shows a decreasing trendline of prior ME WPT receiver $FOM_P$ with frequency. The device presented in this work is a clear outlier from this trendline, with its $FOM_P$ being greater than any comparable-frequency device by a factor of 20[53]. While ME WPT devices of similar size have been reported, those operating at GHz frequencies[44] face much higher electromagnetic tissue losses and lower permissible magnetic field strengths than at 100 kHz, and others near 100 kHz report only ME coupling coefficients without demonstration of power delivery to a matched load[42], precluding direct comparison on the basis of power figure of merit or power density.

## Discussion

The MEMS ME WPT receiver demonstrated in this work integrates AlScN piezoelectric and $Fe_{50}Co_{50}$/Ag ML magnetostrictive thin-films in a heterostructure within a total device volume of $6.24 \times 10^{-4}$ $mm^3$ and delivers 1 μW of power to a matched resistive load at the 205 $\mu T_{RMS}$ whole-body magnetic field limit of IEEE C95.1-2019 at 100 kHz[33], the optimal frequency for ME WPT under human exposure safety constraints.

The device achieves a $FOM_P$ of $4.05 \times 10^{-3}$ mW $cm^{-3}$ $Oe^{-2}$ $Hz^{-1}$, the highest reported among all ME WPT receivers and 20 times greater than the next best device operating within an order of magnitude in frequency[13, 53, 57-60]. Additionally, the power density is the highest of any ME WPT receiver in literature by a factor of more than 50[53, 59], and the ME coupling coefficient of 1274 V $cm^{-1}$ $Oe^{-1}$ exceeds that of similarly sized, comparable-frequency ME WPT devices by more than 50%[40, 42]. These records arise from the combination of the enhanced piezoelectric coefficient of AlScN, the high piezomagnetic coefficient of the $Fe_{50}Co_{50}$/Ag ML magnetostrictive, and the high quality factor achievable through MEMS microfabrication.

The 1 μW delivered by the WPT receiver at the IEEE C95.1-2019 whole-body magnetic field limit is sufficient to power a range of reported implanted bioelectronic devices; neural recorders[26] and amplifier circuitry[52], biomolecule sensors[61], and temperature sensors[62] have been demonstrated at power levels within this range. For devices that require more power but operate in the limbs, such as fingertip force sensors[1], higher magnetic field strengths are permitted, up to 37.9 $mT_{RMS}$ at 100 kHz[33]. Devices in the limbs with higher power requirements can benefit both from the higher permissible field levels and the $H_{AC}^2$ relation of power with magnetic field. Additionally, the WPT receiver presented is highly scalable, with arrays of resonators in series or parallel increasing the power through an increase in output current or voltage respectively. An increase in parallel resonant beams on a single device may also increase the output power more rapidly than the area scaling would suggest. The surface stress COMSOL simulation in Figure 2a shows the resonant beams closest to the tethers experiencing less stress than those at the center. As such, tether losses with additional parallel beams may become a less dominant mechanical loss mechanism, resulting in the power output increasing at a faster rate than the area scaling for additional parallel resonant beams.

The ME WPT receivers demonstrated in this work were tested under 4 mBar vacuum to reduce viscous damping and achieve a high quality factor flexural-mode resonance. Vacuum packaging of the device through thin-film hermetic die-to-die bonding can provide a path toward maintaining this performance in implanted systems[63]. Additionally, a 4 mT magnetic field bias was used to maximize the magnetostriction response of the $Fe_{50}Co_{50}$/Ag ML film. In an implanted system, the DC bias requirement necessitates permanent magnets or electromagnets. While the receiver can be used at a lower DC magnetic field, as low as no bias field at all, the small-signal magnetostriction response, and thus output power, are impacted by the lower bias field. Previous work has shown that a self-biased ME composite is possible through either an exchange bias effect, through particular selection of magnetic and antiferromagnetic materials, or through control of the magnetostrictive material's thin-film residual stresses[56]. Through the self-biasing of the magnetostrictive film, an AC magnetic field coil can produce the bias field and AC excitation field, instead of requiring both an AC coil and on-body permanent magnets or a large electromagnet to produce the bias field.

This work demonstrates a sub-millimeter MEMS ME WPT receiver delivering 1 µW at the IEEE C95.1-2019 whole-body magnetic field safety limit and exhibits a record figure of merit and ME coupling coefficient among comparable-frequency ME WPT devices. Additionally, the power density is the highest reported of ME WPT receivers in literature. The integration of aluminum scandium nitride piezoelectric and iron-cobalt/silver multilayer magnetostrictive thin-films within a microfabricated MEMS heterostructure achieves high magnetoelectric coupling with a high quality factor resonator. Miniaturized magnetoelectric wireless power transfer receivers overcome limitations to sub-millimeter size-scaling that hamper ultrasonic and inductive wireless power modalities and enable the powering of biomedical sensors and systems implanted into anatomically constrained locations, including in the brain, fingertips, and eyes. As implantable bioelectronic devices continue to miniaturize, MEMS magnetoelectric transduction establishes a scalable platform for battery-free wireless power delivery that will allow for the next generation of implantable sensing and stimulation systems.

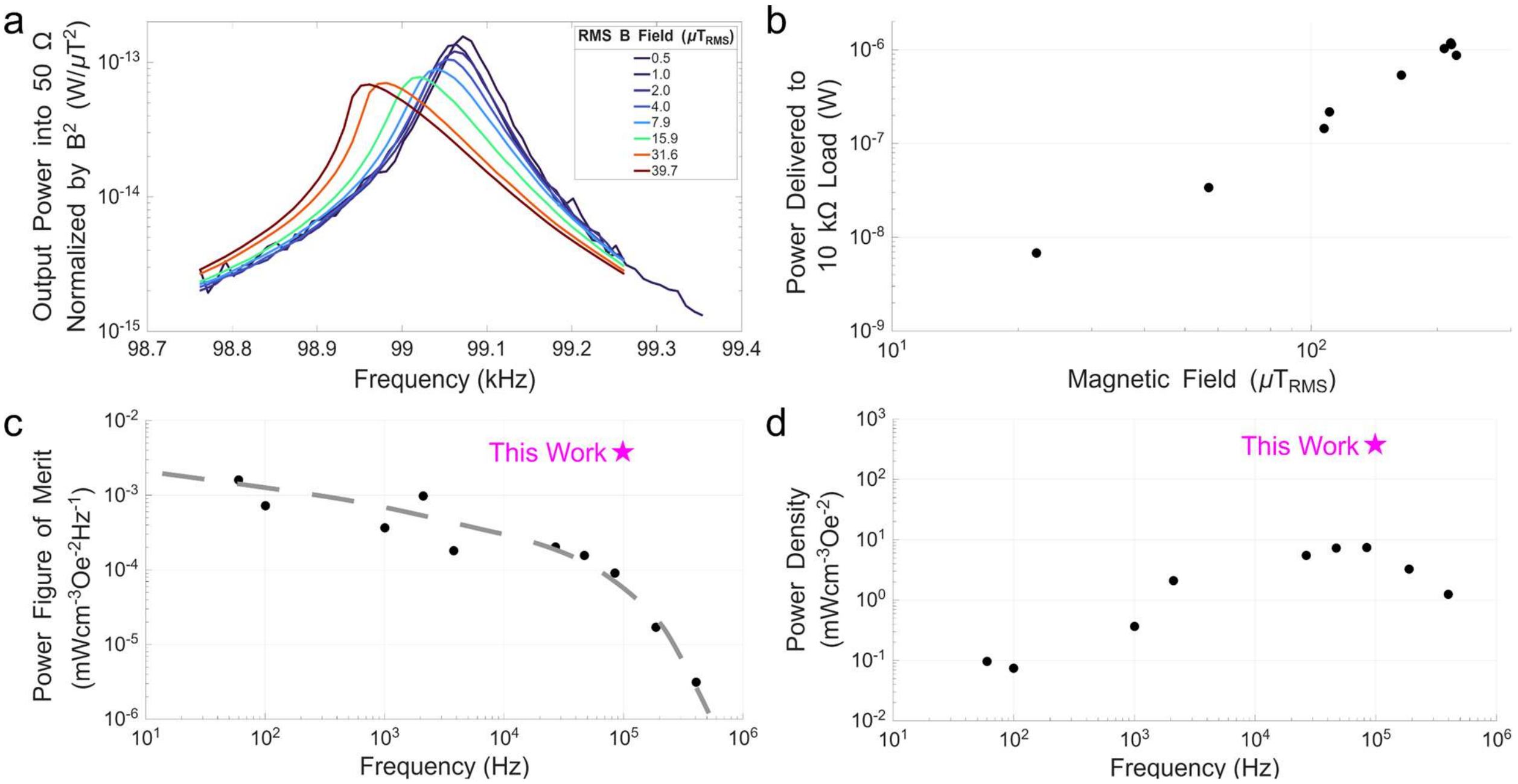

Figure 3. Response with increasing magnetic field and comparison to literature a) Output power spectrum normalized by squared magnetic field into a 50 Ω spectrum analyzer with increasing magnetic field strength, showing Duffing nonlinearity of the resonator at higher excitation magnetic field strengths. b) log-log plot of output power of WPT receiver into a matched 10 kΩ load with increasing magnetic field strength. c) Comparison of figure of merit ($FOM_p$) with various ME WPT receivers in literature, with a trendline of prior $FOM_p$ overlaid[13, 57-60, 64-68]. The MEMS ME WPT receiver presented in this work (pink star) exhibits the highest $FOM_p$, greater than 20 times larger than any other device operating near 100 kHz, the optimum field level for magnetically delivering power into the body. d) Comparison of power density with various ME WPT receivers in the literature[13, 57-60, 64-67]. The ME WPT receiver presented in this work (pink star) demonstrates the highest power density reported of ME WPT receivers in the literature.

Methods

Fabrication

An abbreviated fabrication process flow is described here, with the full fabrication flow detailed in Supplementary Note 5. Devices were fabricated on a 4-inch diameter intrinsic (100) silicon wafer substrate with resistivity of >10 kΩ-cm. A bottom electrode of 15 nm titanium / 150 nm platinum was sputter deposited (Evatec Clusterline 200 II) to produce (111) oriented Pt. An aqua regia etch was masked with S1813 photoresist (Kayaku Advanced Materials Inc.) to etch through the field area of the bottom electrode platinum. The S1813 masking photoresist was stripped with Microposit 1165 remover (Kayaku Advanced Materials Inc.) and oxygen plasma (PIE Scientific Tergeo-Plus Plasma Cleaner).

A 1 μm thick $Al_{0.67}Sc_{0.33}N$ piezoelectric film was pulsed-DC co-sputtered in the same Evatec deposition tool. The AlScN film was grown as a four-layer stack: an AlN seed layer, and AlN-to-AlScN gradient layer, and a bilayer of tensile and compressive AlScN films deposited at 30 standard cubic centimeters per minute (sccm) and 21 sccm $N_2$ respectively, following prior work on stress-neutral AlScN films[69]. A full description of deposition parameters, including powers, deposition times, and target-to-substrate distance are described in Supplementary Note 5, step E.

Vias through the AlScN to the bottom electrode Pt were formed using a silicon nitride hard mask grown by plasma-enhanced chemical vapor deposition (Oxford PlasmaLab 100 PECVD) patterned with S1813 photoresist and etched in a reactive ion etcher (RIE, Oxford 80 Plus RIE). The photoresist was stripped in 1165 remover before a heated potassium hydroxide solution etched through the AlScN. The remaining silicon nitride was stripped with $CF_4$ RIE plasma.

The top electrode was next formed. A silicon dioxide film was deposited using PECVD, and LOR 10A (Kayaku Advanced Materials Inc.) and S1813 photoresists were spun, exposed, and developed. RIE using $CF_4$ plasma was used to etch through the oxide layer and land on the AlScN. A 10 nm Ti / 100 nm Pt top electrode was sputter deposited (Denton Explorer14 Magnetron Sputterer). The top electrode was lifted off in 1165 remover, and the remaining silicon dioxide was stripped in $CF_4$ RIE plasma.

The resonator geometry was created with an etch through the AlScN. A 3 μm silicon dioxide film was deposited by PECVD to act as the etch mask for the AlScN etch. 10 nm Aluminum was deposited above the oxide in the Denton sputterer. S1818 photoresist (Kayaku Advanced Materials Inc.) was spun, exposed, and developed, which additionally etched through the 10 nm Al film. The silicon dioxide was etched through to the AlScN using $CF_4$ RIE plasma with the S1818 mask. Any remaining photoresist was stripped in 1165 remover. A variable-angle ion mill etch described in step K in Supplementary Note 5 was performed to produce a vertical sidewall on the AlScN. Remaining silicon dioxide was stripped with $CHF_3/O_2$ plasma in the RIE.

The $Fe_{50}Co_{50}$/Ag magnetostrictive multilayer was patterned and deposited next. LOR 20B (Kayaku Advanced Materials Inc.) and S1813 photoresists were spun, exposed, and developed. $Fe_{50}Co_{50}$/Ag ML, (10 nm/3 nm)×77, was deposited, with deposition conditions described in step M in Supplementary Note 5. The magnetostrictive film was lifted off in 1165 remover with the assistance of a permanent magnet.

The 4-inch wafer was diced into the 2 × 2 mm die (ADT 7100 Dicing Saw) and released in $XeF_2$ gas (SPTS/Xactix). The released die were die-attached to a breakout PCB with epoxy and wire bonded (Kulicke & Soffa Wire Bonder). Through-hole SMA PCB adapters were soldered to the PCB.

X-ray diffraction

X-ray diffraction rocking curve (RC) scans were measured on the platinum bottom electrode and the AlScN piezoelectric thin films (Rigaku Smartlab SE). Platinum (111) RC was measured at 2θ = 40°, ω = 20°, exhibiting a full width half maximum (FWHM) of 1.65°, and AlScN (002) RC was measured at 2θ = 36°, ω = 18°, exhibiting a FWHM of 1.12°. FWHM measurements of both Pt and AlScN indicate highly oriented thin films. XRD rocking curve scans are shown in Supplementary Note 6.

Test setup

An image and drawing of the test setup are shown in Supplementary Note 7. An acrylic vacuum chamber (Sanatron) housed the device, AC magnetic field coils, and DC magnetic field electromagnets. Copper litz wire and magnet wire (Remington industries) were used to wrap the AC and DC coils respectively on custom 3D-printed coils holders. The current-to-magnetic field response of both AC and DC coils were characterized using a current probe and oscilloscope (Keysight N7026A, Keysight InfiniiVision MSOX3054G) and magnetic field probe (Vernier Go Direct® 3-Axis Magnetic Field Sensor). With the coils and device in the acrylic chamber, a vacuum of 4 mBar was maintained within the chamber using a turbomolecular pumping system (Pfeiffer HiCUBE 80 Eco).

On the two ports of the device, one port was connected to an SMA short standard, and the other port was connected to an SMA to BNC cable. BNC interconnects through the acrylic vacuum chamber were used to supply AC current for the AC coils, DC current for the bias magnetic field, and to measure the output of the device. The AC coils were driven using a low-output-impedance amplifier (Texas Instruments TPA3116D2) and current was supplied for the DC coils with an Agilent E3631A, operated in current controlled mode. A RIGOL

RSA3045 was used to measure the power at a 50 Ω impedance, and a Stanford Research Systems SR770 spectrum analyzer with input impedance of 1 MΩ in parallel with a 15 pF capacitance was used for measurements into a high impedance and for matched load measurements.

For wireless power transfer measurements, the setup described above with the SR770 spectrum analyzer was used, with a 10 kΩ resistor connected between the signal and ground ports of the WPT receiver. The circuit model of the WPT test is shown in Supplementary Note 8, showing the transduction of energy from magnetic to mechanical to electrical domains.

For $S_{21}$ S-parameter measurements, the two ports of the device were connected to an Agilent Technologies E5061B network analyzer, with the input power set to the network analyzer's minimum of -45 dBm to ensure linear operation. Measurements were also performed under 4 mBar vacuum.

## Data Availability

All data supporting the findings of this study are available within the article and its supplementary files. Any additional requests for information can be directed to, and will be fulfilled by, the corresponding author.

Author Contributions

JT: Conceptualization, Methodology, Investigation, Formal Analysis, Validation, Visualization, Writing - Original Draft, Writing - Review and Editing

SF: Investigation, Formal Analysis, Validation, Visualization, Writing - Review and Editing

TM: Methodology, Investigation, Writing - Review and Editing

SA: Conceptualization, Methodology, Formal Analysis, Writing - Review and Editing

KB: Methodology, Conceptualization, Supervision, Funding Acquisition, Writing - Review and Editing, Resources

PF: Methodology, Supervision, Funding Acquisition, Writing - Review and Editing, Resources

RO: Conceptualization, Methodology, Writing - Review and Editing, Funding Acquisition, Supervision, Resources

Competing Interests

The authors declare no competing interests.

Funding Statement

This material is based upon work supported by the National Science Foundation Graduate Research Fellowship Program under Grant No DGE-2236662. Any opinions, findings, and conclusions or recommendations expressed in this material are those of the author(s) and do not necessarily reflect the views of the National Science Foundation.

This work was supported in part by the U.S. Department of Energy, Advanced Research Projects Agency–Energy (ARPA-E), under Award DE-AR0001820.

The work performed at NRL was covered under Office of the Under Secretary of War for Research and Engineering (OUSW R&E) under the Emerging Research Opportunities (ERO) project and the Office of Naval Research (ONR).

This work was performed at the Singh Center for Nanotechnology at the University of Pennsylvania, a member of the National Nanotechnology Coordinated Infrastructure (NNCI) network, which is supported by the National Science Foundation (Grant NNCI-1542153) and through the use of facilities supported by the Laboratory for Research on the Structure of Matter and the NSF through the University of Pennsylvania Materials Research Science and Engineering Center (MRSEC) DMR-2309043.

# Magnetoelectric Wireless Power Transfer Receivers for Sub-Millimeter Scale Implantable Bioelectronics

## Supplementary Information

Authors: Jonathan Tan[1], Serene Feng[1], Thomas Mion[2], Sydney Acosta[1], Konrad Bussmann[2], Peter Finkel[2], Roy Olsson[1*]

[1]Department of Electrical and Systems Engineering, University of Pennsylvania, Philadelphia, PA, USA

[2]Materials Science and Technology Division, US Naval Research Laboratory, Washington, DC 02375, USA

*Email: rolsson@seas.upenn.edu

# Table of Contents

**Supplementary Note 1: Maximum Available Power Calculated at $B^2f$**

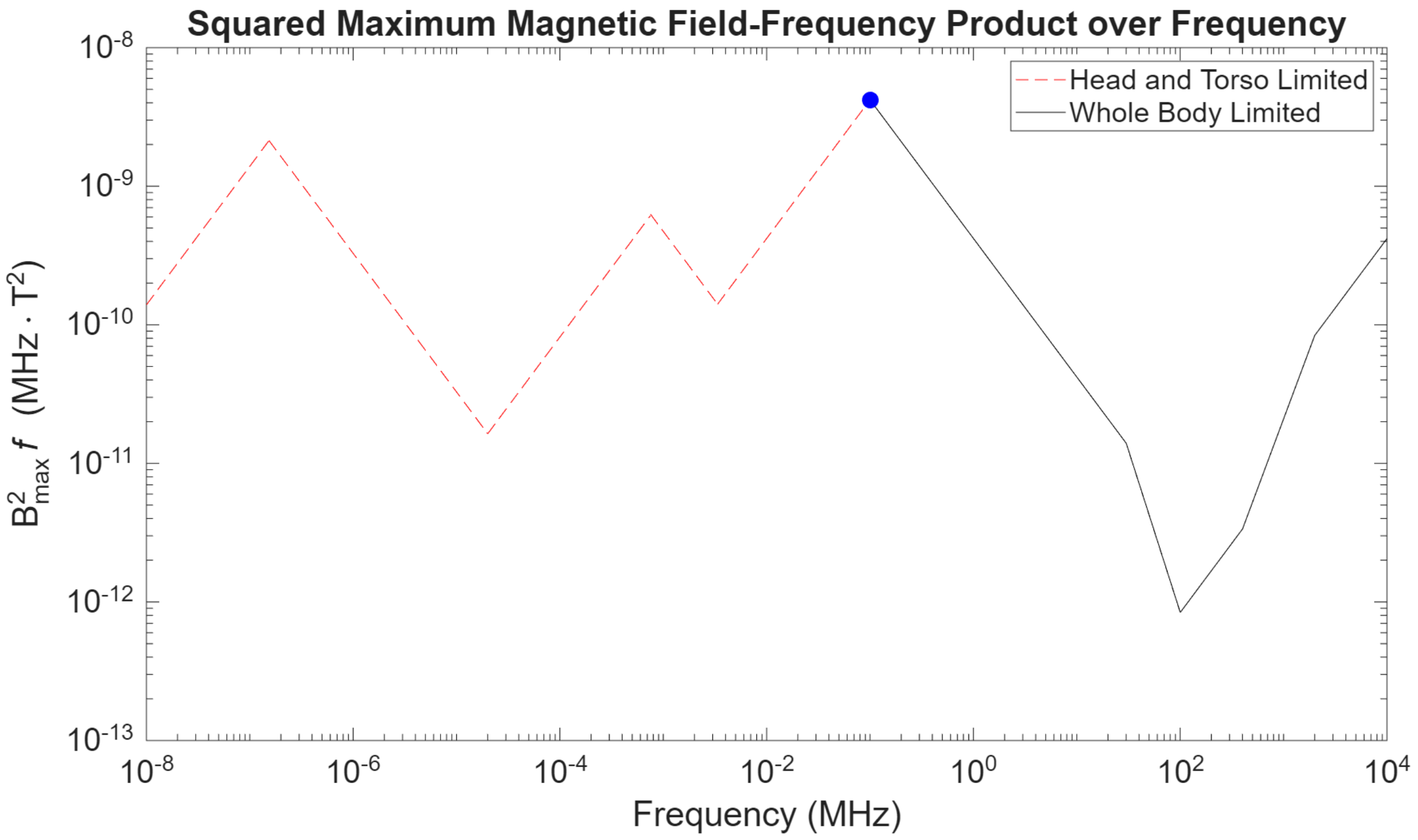

Supplementary Figure 1. The product of frequency and squared magnetic field limit for the head and torso (0.01 Hz to 100 kHz) and whole body (100 kHz to 10 GHz) plotted over frequency. The maximum of the product at 100 kHz is shown in the blue dot.

The product of the squared magnetic field limit over the entire body and frequency is plotted in Supplementary Figure 1, with the frequency range over which the magnetic field strength is limited by the permissible field strength at the head and torso shown in red and the frequency range limited by the permissible field strength across the whole body in black. The figure shows a local maximum of the product at 100 kHz. As magnetoelectric wireless power transfer receiver output power increases linearly with frequency and the square of the magnetic field, 100 kHz is an optimal frequency to design ME WPT receivers.

Limits of the magnetic field strength were sourced from IEEE Std C95.1™-2019[1]. The maximum permissible magnetic field across the whole body in unrestricted environments was limited by either the head and torso (below 100 kHz) or the whole body (above 100 kHz).

**Supplementary Note 2: Modified Butterworth-Van Dyke (mBVD) Model of Resonator**

A modified Butterworth-Van Dyke model was created with motional capacitance, resistance, and inductance extracted from the $S_{21}$ measurement of the WPT receiver at 100 kHz[2, 3]. Motional capacitance, resistance, and inductance were found to be 143 fF, 9.87 kΩ, and 18.1 H respectively. The electrical parallel capacitance of the device was also found through the fitting of the mBVD model to the $S_{21}$ measurement and found to be 28 pF. The mBVD model extracted from the measurement is shown in Supplementary Figure 2a, with the measured and fit $S_{21}$ in Supplementary Figure 2b

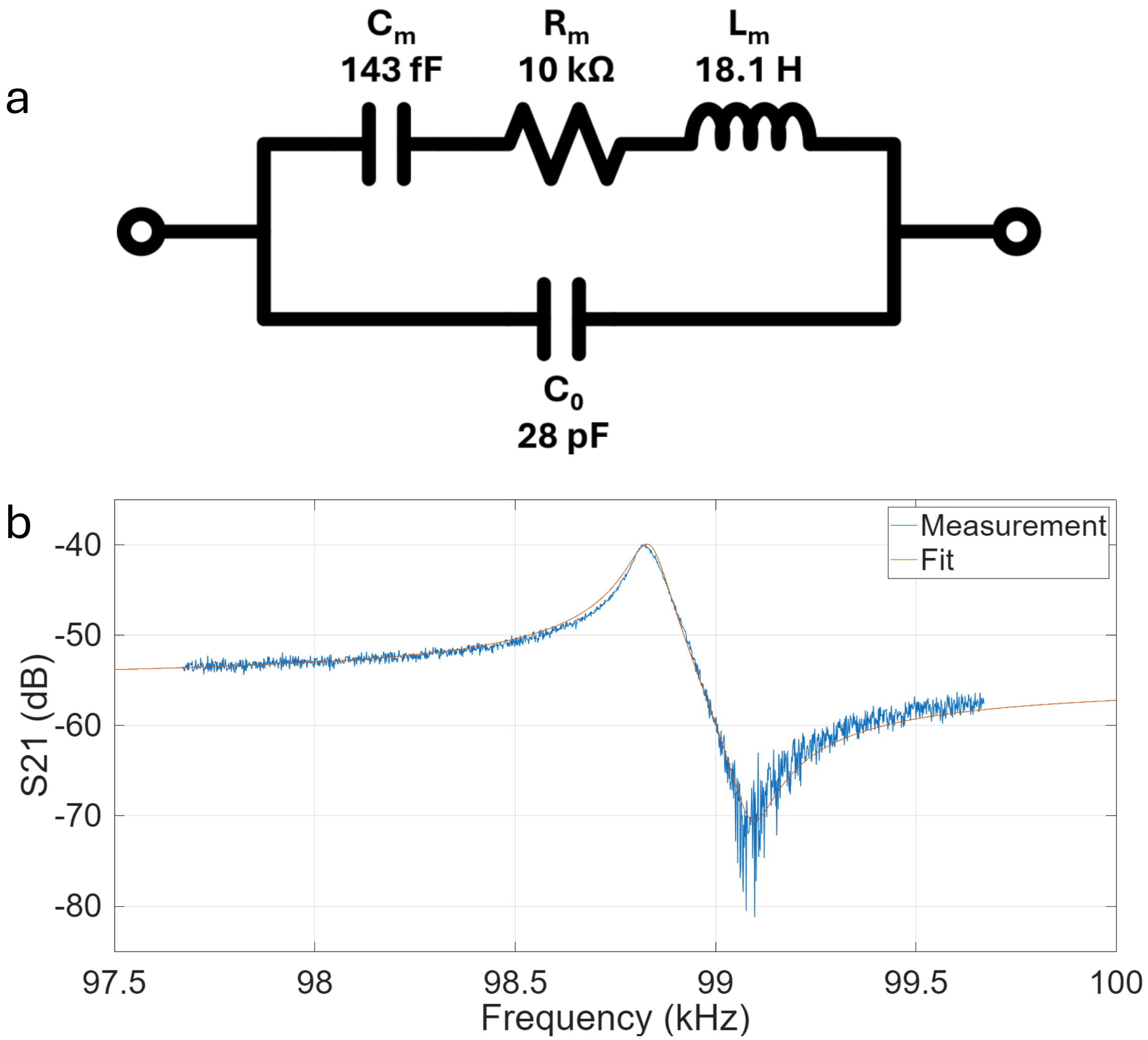

Supplementary Figure 2. a) mBVD model of the resonator at 100 kHz with the fitted motional components and electrical capacitance. b) The measured (blue) and fit (orange) $S_{21}$ transmission S-parameter data, showing a good match between the measured and fit data.

**Supplementary Note 3: Motional Resistance Derivation**

The circuit model describing the S-parameter testing of the WPT receiver is shown below, with ports 1 and 2 of the vector network analyzer shown.

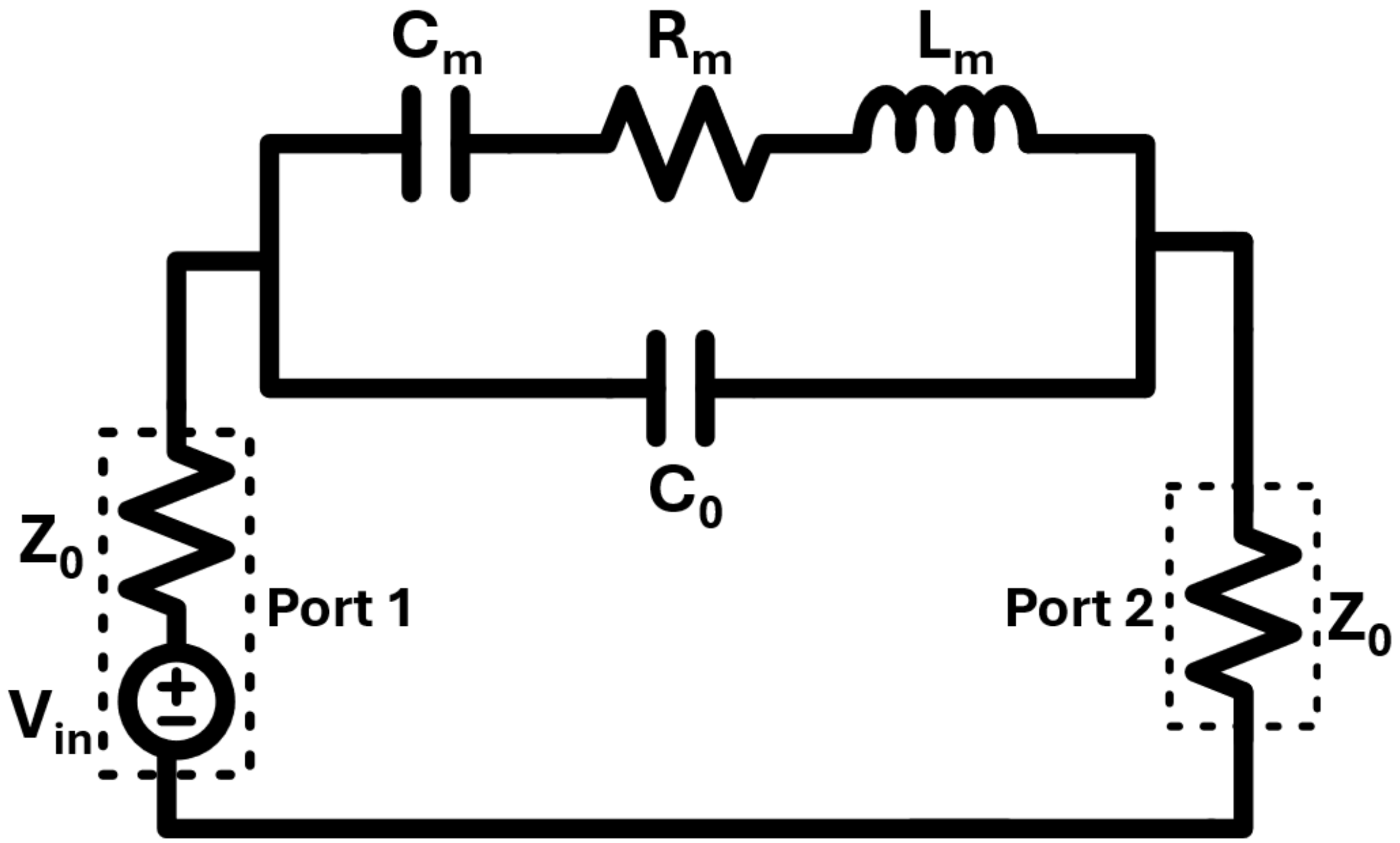

Supplementary Figure 3: The mBVD model of the resonator connected electrically to two ports of a vector network analyzer.

The motional resistance, which is used for determining the resistance required for a matched load, is calculated from the $S_{21}$ S-parameter measurement across the resonator. A circuit diagram of the S-parameter measurement is shown in Supplementary Figure 3. The $S_{21}$ magnitude for a 2-port series impedance is defined as the following, with $Z_0$ being the reference impedance of the network analyzer and $Z_{Series}$ being the series device under test impedance

$$S_{21} = \frac{2Z_0}{Z_{Series} + 2Z_0}$$

If the series impedance is much larger than $2Z_0$, the $S_{21}$ can be approximated as

$$S_{21} \approx \frac{2Z_0}{Z_{Series}}$$

$$Z_{Series} \approx \frac{2Z_0}{S_{21}}$$

Because $S_{21}$ is a measure of reflected voltage, when converting the $S_{21}$ from dB to magnitude, it must be divided by 20

$$S_{21} = 10^{\frac{S_{21,dB}}{20}}$$

Substituting into the expression for the series impedance

$$Z_{Series} \approx \frac{2Z_0}{10^{\frac{S_{21,dB}}{20}}}$$

Using a network analyzer with reference impedance, $Z_0$, of 50 Ω, and measuring at the device resonance, where the reactances of the motional capacitance and inductance are equal and opposite, the device impedance is the motional resistance with value:

$$R_m \approx \frac{100}{10^{\frac{S_{21,dB}}{20}}}$$

For an $S_{21}$ measurement of -39.89 dB, the motional resistance is calculated to be 9.875 kΩ.

## Supplementary Note 4: Duffing Nonlinearity at High Magnetic Field Strength

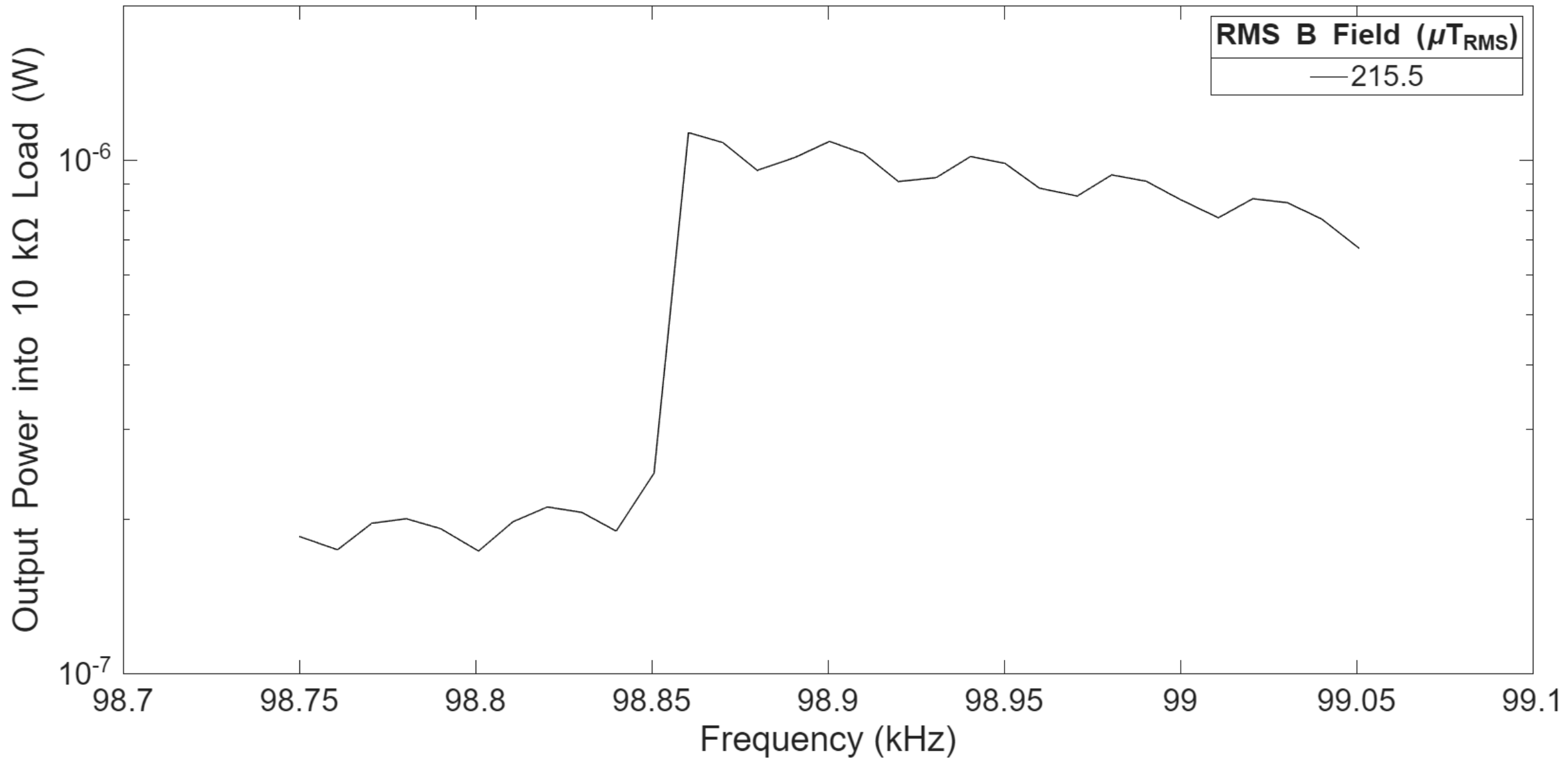

Supplementary Figure 4. Frequency sweep of power output into a 10 kΩ resistive load with magnetic field strength of 215.5 $\mu T_{RMS}$. A discontinuity in frequency of the output power is caused by Duffing nonlinearity.

Large displacement from the high magnetic field causes Duffing nonlinearity in the resonator, observed as spring softening of the resonator. This creates a discontinuity in the output power spectrum, shown in Supplementary Figure 4. Minor amplitude fluctuations visible in the output power spectrum are attributed to scalloping loss, a measurement artifact caused by the windowing of the spectrum analyzer's fast Fourier transform processing using a Blackman-Harris window function.

**Supplementary Note 5: Microfabrication Flow**

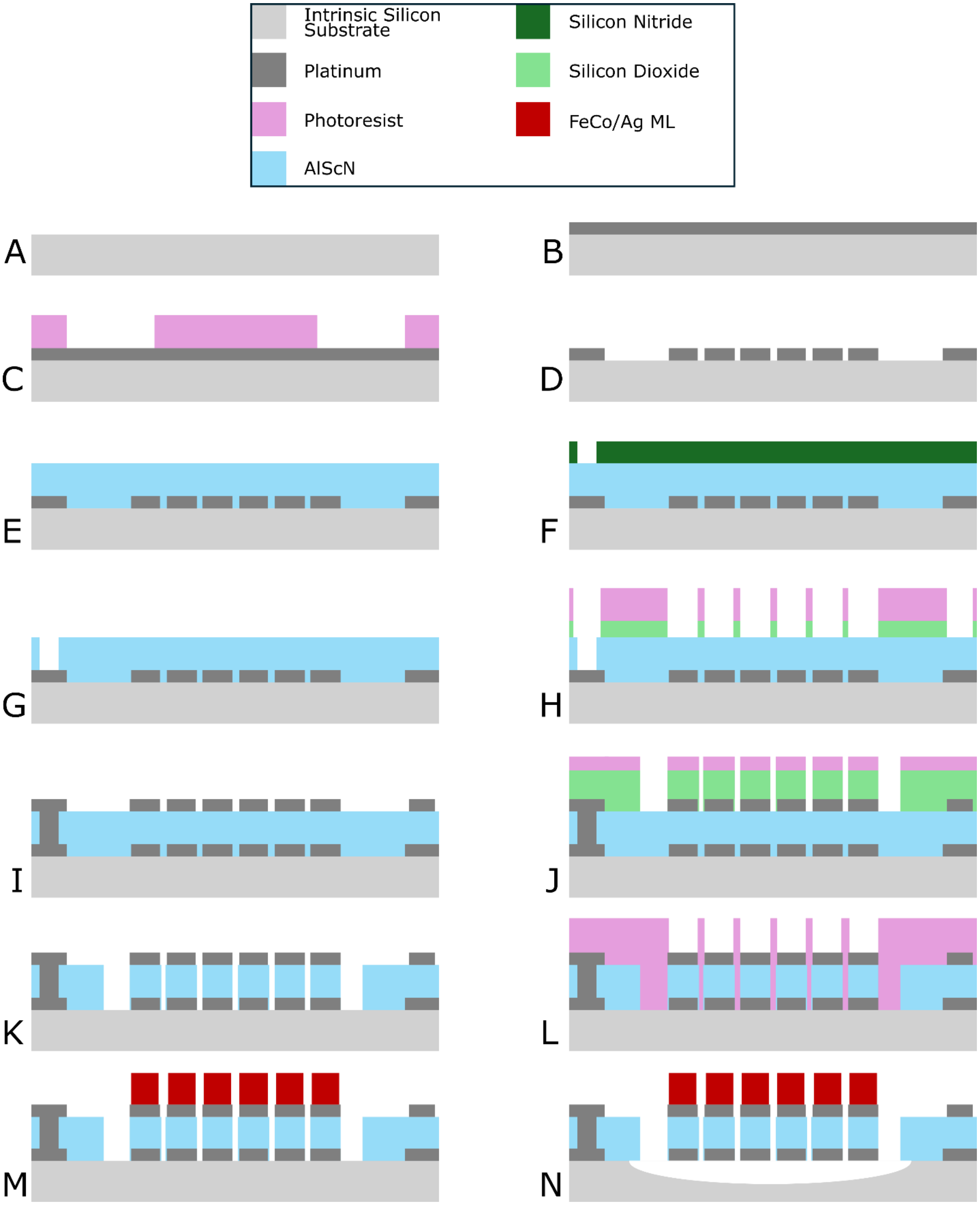

Supplementary Figure 5: A microfabrication flow drawing showing the cross section of the resonator device.

A full fabrication flow is described below, with drawings of the cross section at each step shown in Supplementary Figure 5.

A) A 4-inch diameter intrinsic (100) silicon wafer substrate with resistivity of >10 kΩ-cm was used for device fabrication.

B) Blanket 15 nm titanium and 150 nm platinum were sputtered to produce (111) oriented Pt (Evatec Clusterline 200 II). The chuck temperature was 500 ˚C. Ti was sputtered with DC power (1 kW) for 14 seconds, with a target-to-substrate distance of 101 mm and 45 sccm of argon gas flow into the chamber during deposition. After the Ti deposition, the Pt was DC sputtered (850 W) for 60 seconds, with a target-to-substrate distance of -91 mm and 40 sccm of argon gas flow into the chamber during deposition.

C) S1813 photoresist (Kayaku Advanced Materials Inc.) was used to mask the platinum etch. The photoresist was exposed with 90 $mJ/cm^2$ i-line light (SUSS MicroTec MA6 Gen3 Mask Aligner). MF CD-26 Developer (Microposit) was used for the development of the S1813 photoresist. Unless stated otherwise, all future photoresist exposures and developments use the SUSS MicroTec MA6 and CD-26 developer, respectively.

D) An aqua regia solution of HCl and $HNO_3$ in a volume ratio of 1:4 at 60 ˚C was used to etch the bottom electrode platinum. The S1813 masking photoresist was stripped with Microposit 1165 remover (Kayaku Advanced Materials inc.) and oxygen plasma (PIE Scientific Tergeo-Plus Plasma Cleaner). All photoresist removals use 1165 remover unless stated otherwise.

E) A 1 µm thick aluminum scandium nitride ($Al_{0.67}Sc_{0.33}N$) film was pulsed-DC sputtered in the same Evatec sputter deposition tool. The AlScN thin-film was deposited as four layers: an aluminum nitride (AlN) seed layer, AlN to AlScN gradient, an AlScN film deposited under high gas flow, and an AlScN film deposited under low gas flow, following prior work growing stress-neutral and stress-gradient neutral AlScN films[4]. The AlN seed and gradient provide a gradual alloying of scandium into AlN, and the bilayer AlScN films, controlled by $N_2$ gas flow, allow for an overall stress-neutral and stress-gradient-neutral piezoelectric film stack

For AlScN deposition, the target-to-substrate distance was set to 104 mm, with the chuck temperature at 450 ˚C. An 86 second aluminum nitride seed layer was deposited at 30 sccm $N_2$ gas flow and 1000 W applied to the 100 mm diameter Al target for a 15 nm thickness. For all depositions described for the AlScN stack, the chopper frequency of the pulsed DC sputtering was 150 kHz, with a 800 ns chopper pulse off time. A gradient layer from AlN to AlScN was co-sputtered after the AlN seed layer. The power on the 100 mm diameter scandium target ramped linearly from 0 W to 655 W over the duration of 162 seconds, with the power on the Al target at 1 kW and $N_2$ gas flow at 30 sccm. Once the power on the Sc target reached 655 W, the deposition was continued at the same conditions for an additional 1979 seconds. The 1000/655 W ratio of powers applied to the Al/Sc targets produced a AlScN film with scandium concentration of approximately 33%. After the total 2141 seconds of gradient and AlScN layers was completed at 30 sccm, the $N_2$ gas flow was dropped to 21 sccm, and the deposition was continued at the lower gas flow with the, otherwise, same process conditions for 1980 seconds. The total 4207 seconds of deposition produced a 1 μm AlScN thin-film.

F) Vias were formed through the AlScN to form an electrical contact to the bottom electrode Pt. Non-stoichiometric silicon nitride ($SiN_x$) was grown by plasma-enhanced chemical vapor deposition (Oxford PlasmaLab 100 PECVD), patterned with S1813 photoresist, and etched with $CF_4$ plasma in a reactive ion etcher (RIE) (Oxford 80 Plus RIE). The photoresist was stripped in 1165 remover.

G) The wafer was submersed in 45 °C heated potassium hydroxide (KOH) for about 10 minutes, and vias through the AlScN were etched, masked by the $SiN_x$. The $SiN_x$ mask was stripped with $CF_4$ RIE.

H) To create the top electrode Pt, a 143 nm silicon dioxide ($SiO_2$) layer was grown with PECVD, with a 650 nm LOR 10A (Kayaku Advanced Materials Inc.) and 1.4 μm S1813 spun above the $SiO_2$ to mask the negative of the top electrode area. The $SiO_2$ was etched through with $CHF_3$ and $O_2$ plasma in the RIE.

I) A 10 nm Ti adhesion layer and 100 nm top electrode platinum was sputter deposited (Denton Explorer14 Magnetron Sputterer). The top electrode Pt was lifted off in 1165 remover, and the remaining $SiO_2$ was stripped in $CF_4$ plasma.

J) A 3 μm thick $SiO_2$ mask layer was deposited using PECVD to create a hard mask for AlScN etching. 10 nm of aluminum was sputter deposited above the $SiO_2$ mask (Denton) to provide high RIE etch selectivity. S1818 photoresist (Kayaku Advanced Materials inc.) was spun and patterned. The development of the S1818 photoresist with MF CD-26 Developer (Microposit) additionally etched the 10nm Al layer at the patterned region. The $SiO_2$ mask was etched completely through with $CF_4$ plasma in the RIE. Any remaining photoresist was stripped in 1165 remover.

K) Ion milling through the AlScN was completed using a variable angle etch with accelerating voltage of 500 V. The variable angle etch included a 30 second etch at 10° angle, where a 0° angle equates to the wafer plane being normal to the direction of the ion beam, followed by etching during a 3.7°/second rotation to 70°, and an etch at 70° for 15 seconds. This variable-angle 1 minute etch cycle results in a steep sidewall etch of AlScN, with an AlScN etch rate of approximately 30 nm/minute. The wafer was cooled for 1 minute, and the etch and cooldown cycles were repeated for 40 cycles, resulting in over-etch into the silicon for $XeF_2$ release. After the etch completed, the remaining $SiO_2$ was stripped with $CHF_3/O_2$ RIE.

L) To pattern the magnetostrictive FeCo/Ag ML, LOR 20B (Kayaku Advanced Materials inc.) and S1813 were spun on the wafer and patterned.

M) The 1 µm FeCo/Ag ML was sputter deposited as a 77 multilayer of 10nm $Fe_{0.5}Co_{0.5}$ separated by 3 nm Ag (AJA Internation Inc. Orion). The Ag target was mounted in an unbalanced configuration, and the $Fe_{0.5}Co_{0.5}$ target was mounted in a flux concentrator configuration. The FeCo was deposited with pulsed DC power (100W / 100 kHz, 2 µs), and the Ag was deposited with 50 W RF power. Argon gas was flowed into the system at a rate of 30 sccm, with the chamber pressure held at 3 mTorr during deposition. A DC voltage bias was applied to the sample substrate holder of -95 V during deposition, yielding a nearly stress neutral film (0 +/- 20 MPa). The FeCo/Ag ML was capped with 20 nm of Pt before venting the chamber to prevent oxidation of the magnetic material. After deposition, the FeCo/Ag ML was lifted off in 1165 Remover with the assistance of a permanent magnet hovered over the wafer.

N) For dicing and device release, S1813 photoresist was spun and soft baked. The wafer was diced into 2 mm by 2 mm die (ADT 7100 Dicing Saw). The photoresist was stripped in 1165 remover, and individual devices were released using $XeF_2$ gas (SPTS/Xactix).

**Supplementary Note 6: Rocking curve measurements of Pt and AlS cN**

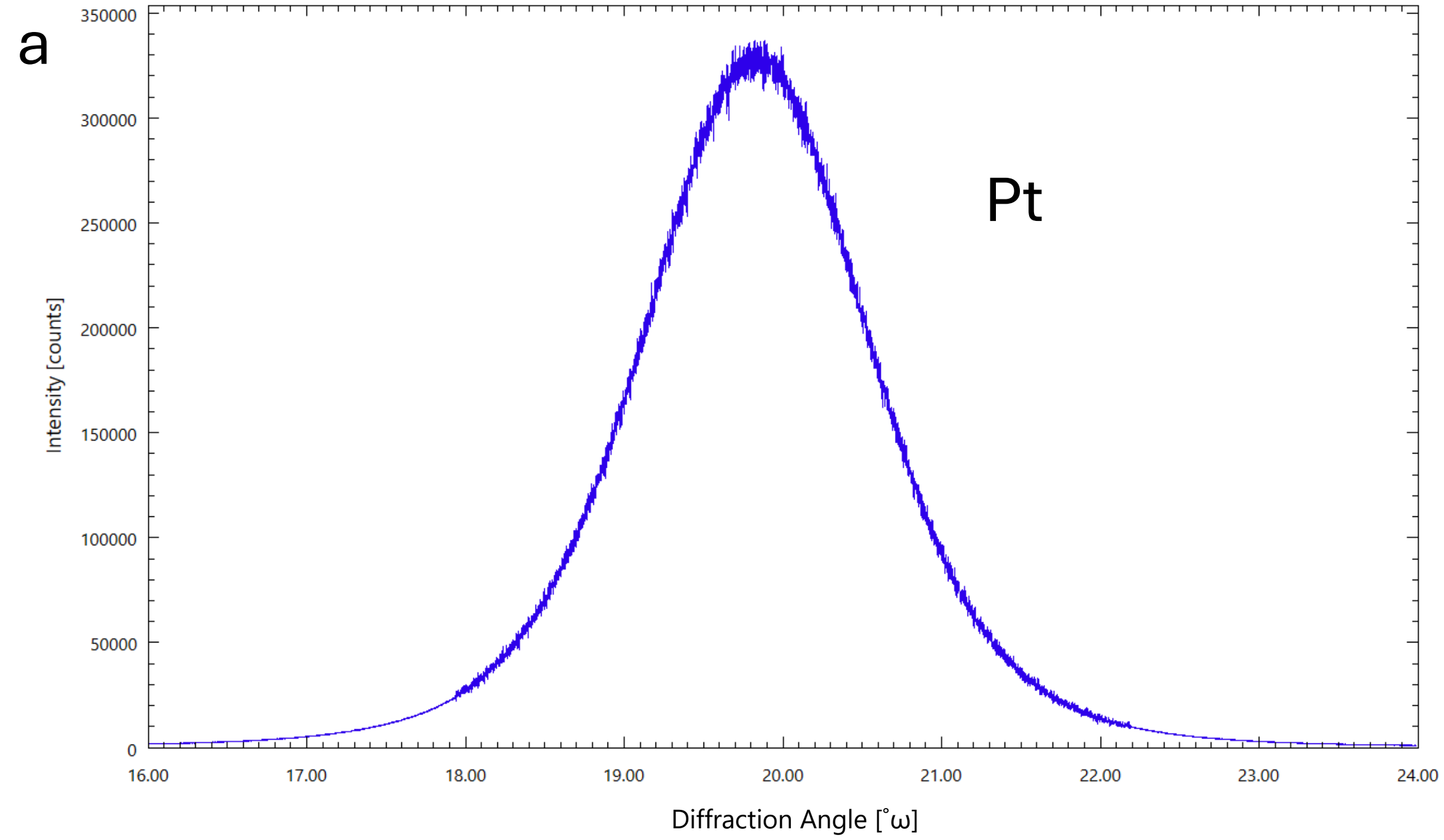

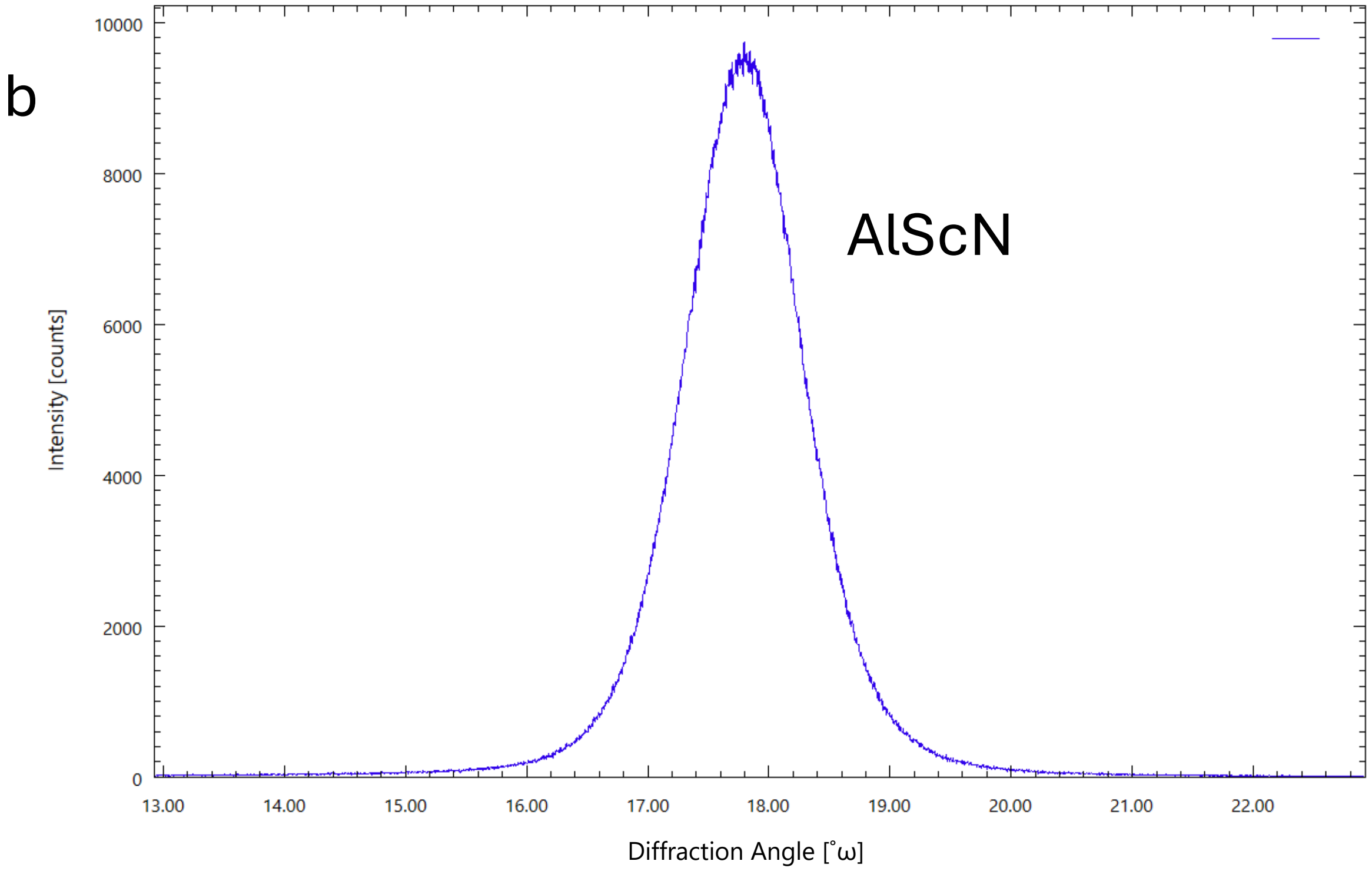

Supplementary Figure 6. a) The rocking curve measurement on the bottom electrode platinum was centered at 19.83˚, with a full-width half-maximum (FWHM) of 1.65˚. b) The rocking curve measurement on the AlScN was centered at 17.79˚, with a FWHM of 1.12˚. The FWHM suggests highly c-axis oriented AlScN.

X-ray diffraction rocking curve (ω) scans on the deposited bottom electrode platinum and piezoelectric aluminum scandium nitride (AlScN) were measured using a Rigaku Smartlab SE, with a Ge(220)×2 monochromator in the incident optics. The rocking curve scans on the Pt and AlScN are shown below in Supplementary Figure 6a and 6b respectively.

**Supplementary Note 7: Test Setup Drawing and Picture**

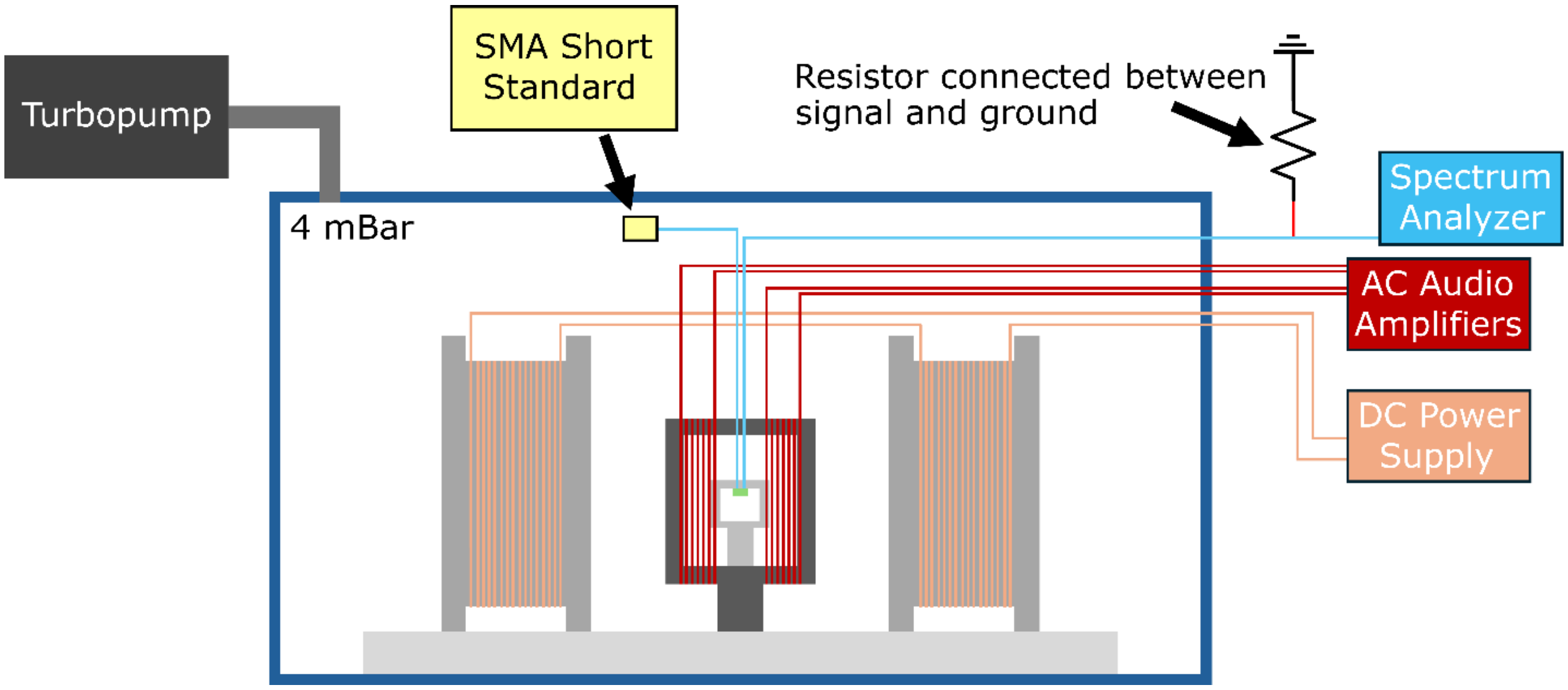

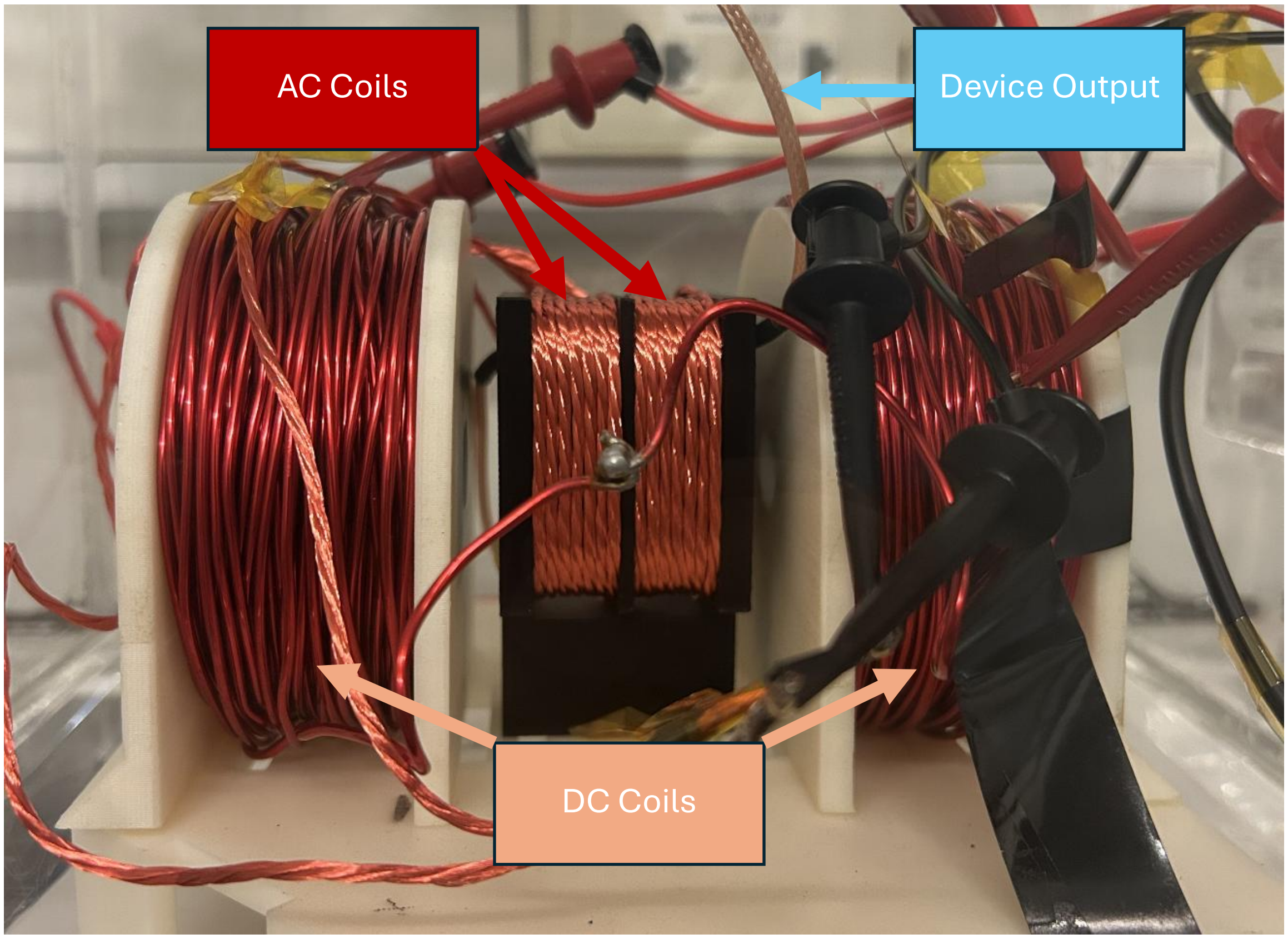

Supplementary Figure 7. a) A drawing of the test setup for wireless power transfer to a resistive load, showing the acrylic chamber pumped to 4 mBar vacuum, a 3D printed enclosure for the device, and AC and DC coils surrounding the device. b) A photograph of the test setup. The DC electromagnet coils flank either side of the black enclosure for the device. The AC Litz wire coils surround the black enclosure. The device is not pictured here, as it is within the black enclosure.

A drawing of the test setup for the WPT receiver is depicted in Supplementary Figure 7a. Specifically, the test setup shown is for wireless power transfer. One port of the device is connected via a short standard to ground, while the other port has a resistor connected between the signal and ground electrodes in parallel with the high input impedance spectrum analyzer. A picture of the test setup is shown in Supplementary Figure 7b.

**Supplementary Note 8: WPT Testing Circuit Model**

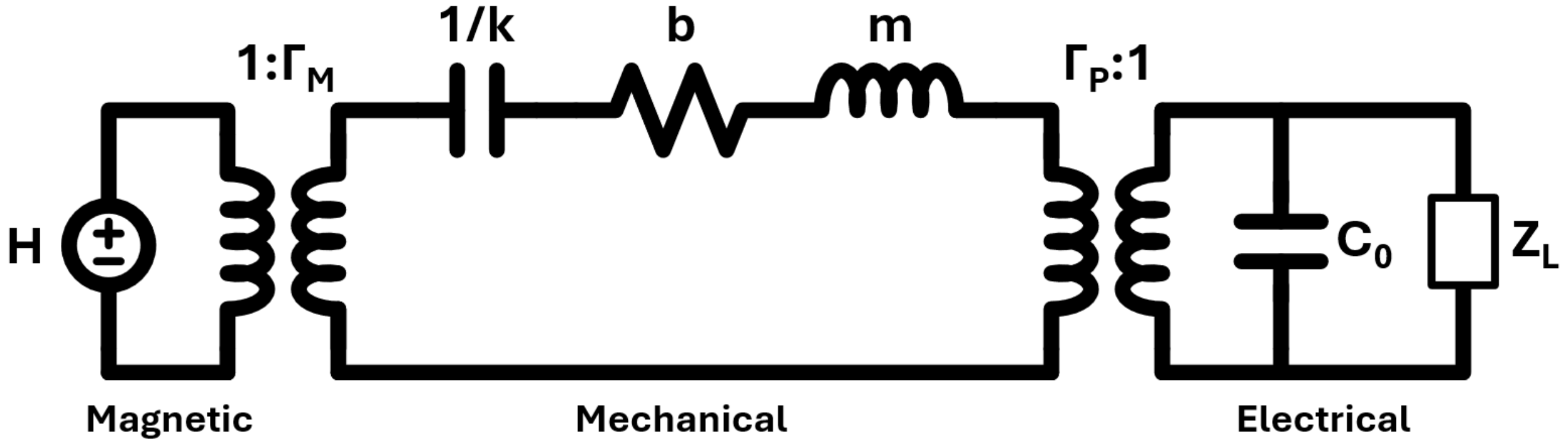

Supplementary Figure 8: Circuit model of the WPT test, including magnetic, mechanical, and electrical domains.

A circuit model of the wireless power transfer testing is shown in Supplementary Figure 8. The magnetic domain is driven by an AC magnetic field source H, representing the magnetic field at the device location generated by the AC magnetic field coils. Magnetic field energy is converted into mechanical energy through the magnetostrictive material, with the energy transduction modeled by the transformer with turns ratio $1:\Gamma_M$. The mechanical lumped circuit components are within the mechanical domain. The capacitor with capacitance 1/k, where k is the spring constant of the mechanical system, represents the potential energy stored in the system. The resistor with resistance b, where b is the damping, considers the loss in the mechanical system. The inductor with inductance m, where m is the effective mass of the mechanical system, represents the kinetic energy stored in the system. Mechanical energy is converted into electrical energy through the piezoelectric material, with the energy transduction modeled by the transformer with turns ratio $\Gamma_P:1$. In the electrical domain, a parallel capacitor with capacitance $C_0$ models the capacitance across the piezoelectric film, and a load with impedance $Z_L$ is connected in parallel across the device output.